\documentclass[12pt]{article}
\usepackage{amsmath,epsfig}
\input{epsf}
\def\ltap{\raisebox{-.4ex}{\rlap{$\,\sim\,$}} \raisebox{.4ex}{$\,<\,$}}

\newcommand\as{\alpha_{\mathrm{S}}}

\def\beeq{\begin{eqnarray}}
\def\eeeq{\end{eqnarray}}

\def\to{\rightarrow}

\begin{document}
\begin{titlepage}
\renewcommand{\thefootnote}{\fnsymbol{footnote}}
\begin{flushright}
CERN-PH-TH/2004-107\\
    hep-ph/0406156
     \end{flushright}
\par \vspace{10mm}
\begin{center}
{\Large \bf
The transverse momentum distribution \\[0.8ex]
of the Higgs boson at the LHC}
\footnote{Talk given at the XXXIXth Rencontres de Moriond, La Thuile, Italy, 28 March--4 April 2004.}

\end{center}
\par \vspace{2mm}
\begin{center}
{\bf Massimiliano Grazzini}\\

\vspace{5mm}

Theory Division, CERN, CH-1211 Geneva 23, Switzerland

\end{center}

\par \vspace{2mm}
\begin{center} {\large \bf Abstract} \end{center}
\begin{quote}
\pretolerance 10000

We present perturbative predictions for the transverse momentum ($q_T$) distribution of the Higgs boson at the LHC. At small $q_T$ the
logarithmically-enhanced terms are resummed to all orders up
to next-to-next-to-leading logarithmic accuracy.
The resummed component
is consistently matched to the next-to-leading
order calculation valid at large $q_T$. The results,
which implement
the most advanced perturbative information that is available at present
for this observable,
show a good stability with respect to perturbative QCD uncertainties.

\end{quote}

\vspace*{\fill}
\begin{flushleft}
     hep-ph/0406156 \\June 2004

\end{flushleft}
\end{titlepage}

\setcounter{footnote}{1}
\renewcommand{\thefootnote}{\fnsymbol{footnote}}

The search for the Higgs boson is among the major issues in the LHC physics
program \cite{atlascms}.
In recent years much effort has been devoted to refining
the theoretical predictions for the various Higgs production
channels and the corresponding backgrounds, which are now known
to next-to-leading order accuracy (NLO) in most of the cases \cite{leshouches}.
For the main
Standard Model (SM) Higgs production channel, gluon--gluon fusion,
even next-to-next-to leading (NNLO) QCD corrections to the
total rate have been computed \cite{NNLOtotal}.
Nonetheless, predictions for less inclusive observables
are definitely required to perform realistic studies.
In particular, an accurate knowledge of the transverse-momentum
distribution of the Higgs boson can be important to enhance the 
statistical significance of the signal over the background.

In this contribution we focus on the dominant SM Higgs production channel,
gluon--gluon fusion.
When the transverse momentum $q_T$ of the Higgs boson is of the
order of its mass
$M_H$, the perturbative series is controlled by a small expansion
parameter, $\as(M_H^2)$, and the fixed order prediction is reliable.
The leading order (LO) calculation \cite{Ellis:1987xu}
shows that the large-$M_t$ 
approximation ($M_t$ being the mass of the top quark) works well as long
as both $M_H$ and $q_T$ are smaller than $M_t$.
In the framework of this approximation, the NLO QCD corrections
have been computed \cite{deFlorian:1999zd,Ravindran:2002dc,Glosser:2002gm}.

The small-$q_T$ region ($q_T\ll M_H$) is the most important, because
it is here that the bulk of events is expected. In this region
the convergence of the fixed-order expansion is spoiled, since
the coefficients of the perturbative series in $\as(M_H^2)$ are enhanced
by powers of large logarithmic terms, $\ln^m (M_H^2/q_T^2)$. To obtain
reliable perturbative predictions, these terms have 
to be systematically resummed to all orders in $\as$ \cite{Dokshitzer:hw}
(see also the list of references in Sect.~5 of
Ref.~\cite{Catani:2000jh}).
To correctly enforce transverse-momentum conservation,
the resummation has to be carried out in $b$ space, where the impact parameter
$b$ is the variable conjugate to $q_T$ through a Fourier transformation.
In the case of the Higgs boson,
the resummation has been explicitly worked out at
leading logarithmic (LL), next-to-leading logarithmic (NLL) 
\cite{Catani:vd}, \cite{Kauffman:cx}
and next-to-next-to-leading logarithmic (NNLL) \cite{deFlorian:2000pr} level.
The fixed-order and resummed approaches then have
to be consistently matched at intermediate values of $q_T$,
so as to avoid double counting.

In the following
we present predictions for the Higgs boson $q_T$ distribution at the LHC
within the formalism described in Ref.~\cite{Catani:2000vq}.
In particular, we include the
best
perturbative information that is available at present:
NNLL resummation at small $q_T$ and NLO calculation at large $q_T$.
An important feature of our formalism is that a unitarity constraint
on the total
cross section is automatically enforced, such that
the integral of the spectrum reproduces the
known results at NLO \cite{Dawson:1991zj} and NNLO \cite{NNLOtotal}.
More details can be found in Ref.~\cite{Bozzi:2003jy}.
Other phenomenological studies are presented in \cite{recent}.

We are going to present quantitative results at NLL+LO and NNLL+NLO
accuracy. 
At NLL+LO (NNLL+NLO) accuracy the NLL (NNLL) resummed result is matched
to the LO (NLO) perturbative calculation valid at large $q_T$. 
As for the evaluation of the fixed-order results, the Monte Carlo program 
of Ref.~\cite{deFlorian:1999zd} has been used.
The numerical results are obtained by choosing $M_H=125$~GeV and using 
the MRST2002 set of parton distributions \cite{Martin:2003es}.
At NLL+LO, LO parton densities and 
1-loop $\as$ have been used, whereas at NNLL+NLO
we use NLO parton densities 
and 2-loop $\as$.
\begin{figure}[htb]
\begin{center}
\begin{tabular}{c}
\epsfxsize=12truecm
\epsffile{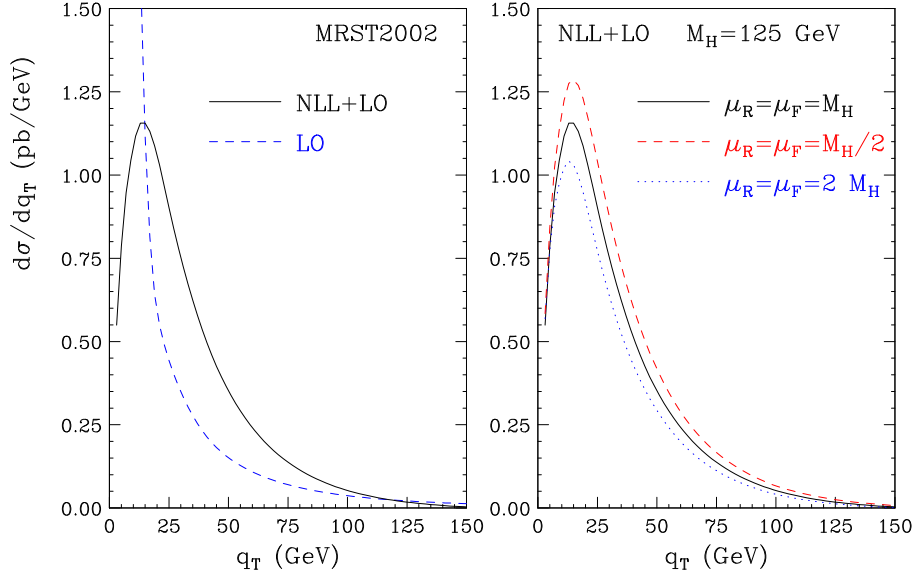}\\
\end{tabular}
\end{center}
\caption{\label{fig1}
{\em 
LHC results at NLL+LO accuracy.}}
\end{figure}
The NLL+LO results at the LHC are shown in Fig.~\ref{fig1}.
In the left panel, the full NLL+LO result (solid line)
is compared with the LO one (dashed line)
at the default scales $\mu_F=\mu_R=M_H$.
We see that the LO calculation diverges to $+\infty$ as $q_T\to 0$. 
The effect of the resummation starts to be relevant below $q_T\sim 100$~GeV.
In the right panel we show the NLL+LO band obtained
by varying $\mu_F=\mu_R$ between $1/2 M_H$ and $2M_H$.
\begin{figure}[htb]
\begin{center}
\begin{tabular}{c}
\epsfxsize=12truecm
\epsffile{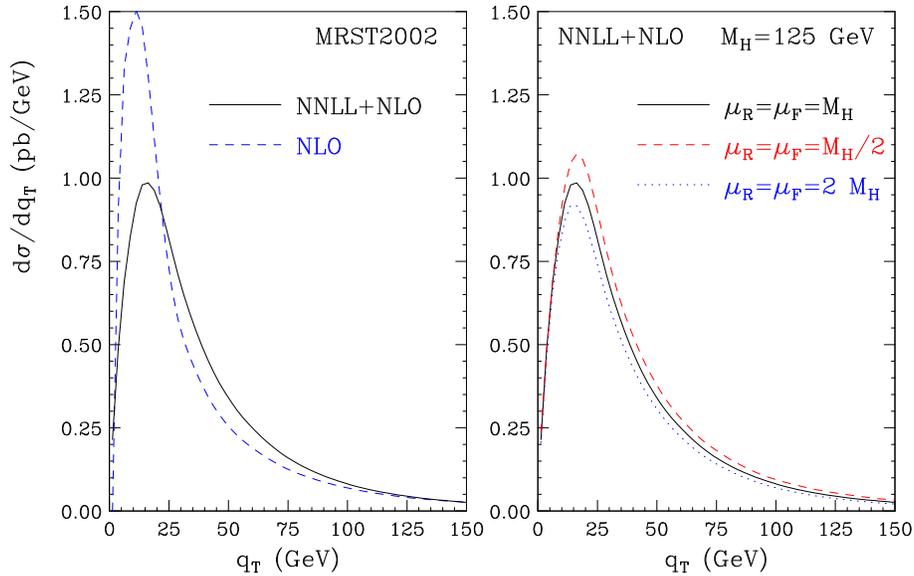}\\
\end{tabular}
\end{center}
\caption{\label{fig2}
{\em 
LHC results at NNLL+NLO accuracy. }}
\end{figure}
The corresponding NNLL+NLO results are shown in Fig.~\ref{fig2}.
In the left panel, the full result (solid line)
is compared with the NLO one (dashed line) at the
default scales $\mu_F=\mu_R=M_H$.
The NLO result diverges to $-\infty$ as $q_T\to 0$ and, at small values of 
$q_T$, it has an unphysical peak
that is produced by the numerical compensation of negative
leading and positive subleading logarithmic contributions.
Notice that at $q_T \sim 50$~GeV, the 
$q_T$ distribution sizeably increases when going from LO to NLO and from NLO
to NLL+LO. This implies that in the intermediate-$q_T$ region there are
important contributions that have to be resummed to all orders rather than
simply evaluated at the next perturbative order.
The resummation effect starts to be visible below $q_T\sim 100$~GeV, and 
it increases the NLO result by about $40\%$ at $q_T=50$~GeV.
The right panel of Fig.~\ref{fig2} shows the scale dependence computed as
in Fig.~\ref{fig1}.
Comparing Figs.~1 and 2, we see that the NNLL+NLO band is smaller 
than the NLL+LO one and overlaps with the latter at $q_T \ltap 100$~GeV.
This suggests a good convergence of the resummed perturbative expansion.

\begin{figure}[htb]
\begin{center}
\begin{tabular}{c}
\epsfxsize=10truecm
\rotatebox{-90}{
\epsffile{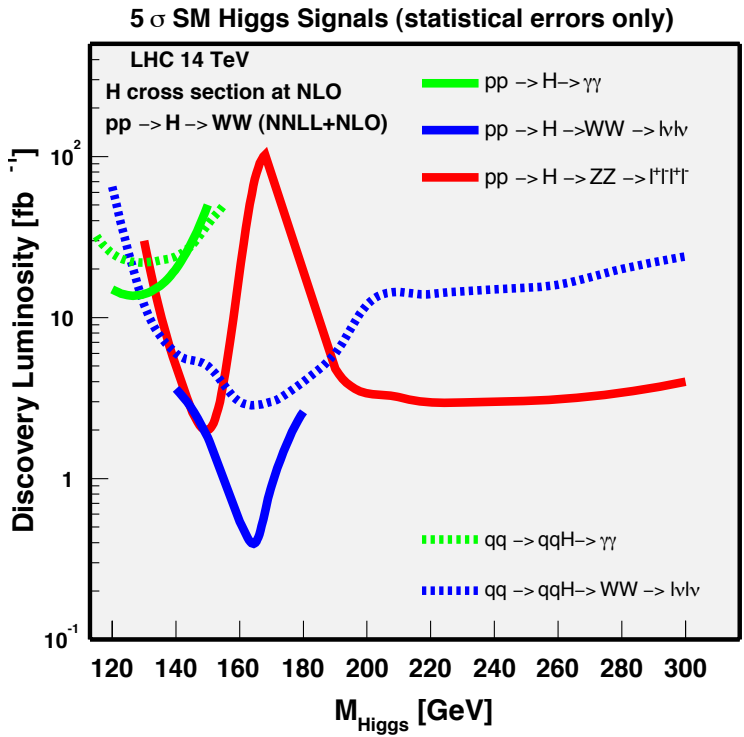}
}
\end{tabular}
\end{center}
\caption{\label{fig3}
{\em 
Minimal Higgs discovery luminosities (from M. Dittmar).}}
\end{figure}
A practical application of the results discussed
in this contribution
has recently been
presented \cite{dddgp} in the context of
the Higgs search in the channel $pp\to H\to WW\to l\nu l\nu$.
In this channel a jet veto is necessary to cut events with high-$q_T$ $b$ jets
from the $t{\bar t}$ background. It is known that the effect
of higher order corrections depends
on the actual value of this cut \cite{Catani:2001cr}.
As an approximate way to include higher order effects in the
analysis,
the QCD-corrected Higgs spectrum
presented above
has been used in Ref.~\cite{dddgp} to reweight
signal events generated with the PYTHIA Monte Carlo \cite{pythia}.
The same method has been applied to correct the main
$WW$ background, by using
a NLL+LO calculation performed within the
same resummation formalism \cite{Catani:2000vq}.

The resulting integrated luminosity needed
to discover the Higgs is shown
in Fig.~\ref{fig3} and compared
with the other channels.
We see that, if systematical errors are under control,
 around the $WW$ threshold the Higgs can
be discovered with an integrated luminosity of about half fb$^{-1}$.

\end{document}